# COMPOSITION OF THE SOLAR INTERIOR: INFORMATION FROM ISOTOPE RATIOS

O. Manuel[1] and Stig Friberg[2]

[1] University of Missouri, Nuclear Chemistry, Rolla, MO 65409 U.S.A.
tel: +1 573-341-4420 / fax: +1 573-341-6033 / e-mail: om@umr.edu
[2] Clarkson University, Physical Chemistry, 641 Spring Hill Estate, Eminence, KY 13699-5814 U.S.A.

ABSTRACT

Measurements are reviewed showing that the interior of the Sun, the inner planets, and ordinary meteorites consist mostly of the same elements: Iron, oxygen, nickel, silicon, magnesium, sulfur and calcium [1]. These results do not support the standard solar model.

INTRODUCTION

Reynolds found the decay product of extinct $^{129}$I and an unusual abundance pattern of the other xenon isotopes in meteorites in 1960 [2,3]. Since $^{129}$I exceeded that expected if the solar system formed from an interstellar cloud, Fowler et al. [4] suggested that D, Li, Be, B and extinct $^{129}$I and $^{107}$Pd might have been produced locally in the early solar system.

In 1965 the decay product of extinct $^{244}$Pu [5], a nuclide that could only be made by rapid neutron-capture in a supernova (SN), was found in meteorites.

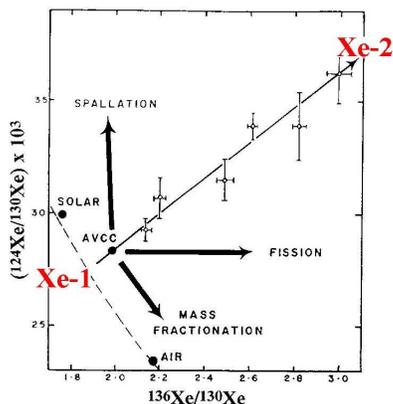

*Figure 1: Normal and "strange" xenon in meteorites.*

In 1972 two major types of xenon were discovered in meteorites [6]: "Normal" xenon (Xe-1) at the lower left of Fig. 1 and "strange" xenon (Xe-2) from a SN at the upper right. Xe-1 includes fractionated forms along a dashed line in the lower left corner. In 1975 primordial helium was shown to have accompanied Xe-2, not Xe-1, at the birth of the solar system [7,8].

This link of primordial helium with Xe-2, shown in Fig. 2 for the Allende meteorite [7,8], was later confirmed in diverse types of meteorites [10,11]. The amount of Xe-2 in the early solar system was sufficient to shift the composition of bulk xenon in meteorites (Point AVCC in Fig. 1) away from Xe-1.

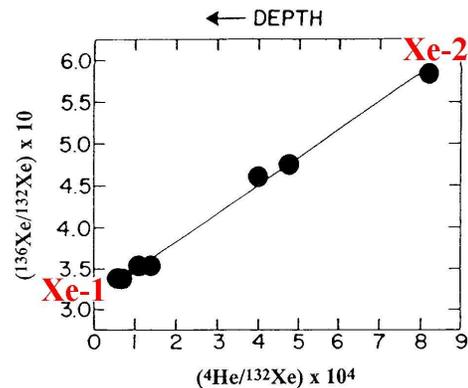

*Figure 2: The link of primordial helium with xenon isotopes in the Allende meteorite [9].*

These results from 1960 to 1975 suggested that local element synthesis produced more elements than Fowler et al. [4] had imagined: Debris from a single supernova (SN) may have formed the entire solar system [7,8]. The Sun formed on the collapsed SN core. Cores of inner planets formed out of iron-rich material in the central region. The outer planets formed out of light weight elements from the outer SN layers. Fig. 3 outlines this scenario [7,8].

Measurements after 1975 confirmed that Xe-1 is linked with iron and that Xe-1 is dominant in the inner solar system. They showed that Xe-2 is linked with light elements and with gaseous planets in the outer solar system. These measurements also yielded affirmative answers to the following questions:

A. Is a supernova the only viable source for Xe-2?
B. Does radioactive age dating indicate a supernova at the birth of the solar system?
C. Are elements in the outer planets different from those in the Sun and the inner planets?
D. Is the interior of the Sun iron-rich, like ordinary meteorites and the inner planets?
E. Can luminosity and solar neutrinos be explained if the Sun formed on a collapsed supernova core?

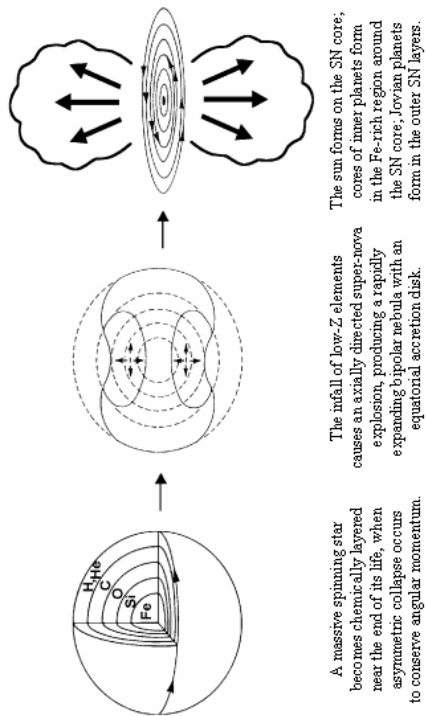

*Figure 3: Birth of the solar system from a supernova*

## POST-1975 RESULTS

**A.** Some proponents of the standard model continued to attribute Xe-2 to super-heavy element fission until the early 1980s [12], but there is now agreement that Xe-2 contains r-products from a supernova rather than fission products. Relic interstellar grains that formed near a supernova did not carry Xe-2 and other elements with excess r-, p- and/or s-products into the early solar system [13]. This would not explain the link of Xe-2 with primordial helium. Further, there is no convincing evidence of interstellar grains that are older than the host meteorites or were irradiated with cosmic rays before being embedded in them. A supernova is now accepted as the only viable source for Xe-2.

**B.** Kuroda and Myers [14,15] combined $^{244}$Pu-$^{136}$Xe and U,Th-Pb dating methods to show the presence of a supernova explosion about 5 Gy ago, at the birth of the solar system. This is shown in Fig. 4. Other laboratories found evidence of a supernova in isotopic anomalies and short-lived nuclides trapped in meteorites. Isotopes of molybdenum remained unmixed, even in massive iron meteorites [16]. Chemical separations occurred before many short-lived r-products had decayed away, within 10,000 sec of the supernova explosion [17,18].

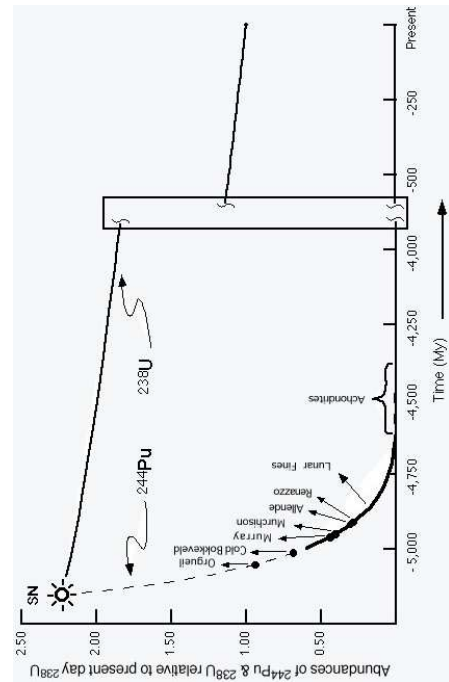

*Figure 4: Five Gy ago a supernova produced $^{244}$Pu.*

**C.** Measurements revealed Xe-1 trapped in troilite (FeS) inclusions of diverse meteorites, including the Allende meteorite [19-22]. Xe-1 is also in Earth and Mars, planets rich in Fe and S. Xe-2 was found to accompany primordial helium in different types of meteorites, including carbonaceous chondrites [10,11]. The *Galileo* mission also found Xe-2 in the helium-rich atmosphere of Jupiter [23,24]. In the solar wind, in Xe-1, in Xe-2, and in Jupiter, ($^{136}$Xe/$^{134}$Xe) = 0.80, 0.84, 1.04, and 1.04 ± 0.06, respectively. Experimental data from the *Galileo* probe into Jupiter can be viewed on the web at <http://www.umr.edu/~om/abstracts2001/windleranalysis.pdf>

**D.** Measurements showed the presence of Xe-1 in the Sun, but light (L) xenon isotopes in the solar wind are enriched relative to the heavy (H) ones by about 3.5% per amu [25]. As shown in Fig. 5, light isotopes of He, Ne, Ar, Kr and Xe in the solar wind all follow a common mass-dependent fractionation power law, where the fractionation factor, **f**, is

$$\log (f) = 4.56 \log (H/L) \qquad (1)$$

Application of Eq. (1) to elemental abundance in the photosphere [25] shows that the seven most abundant elements in the interior of the Sun are the same ones that comprise 99% of meteorites: Iron, nickel, oxygen, silicon, sulfur, magnesium and calcium [1].

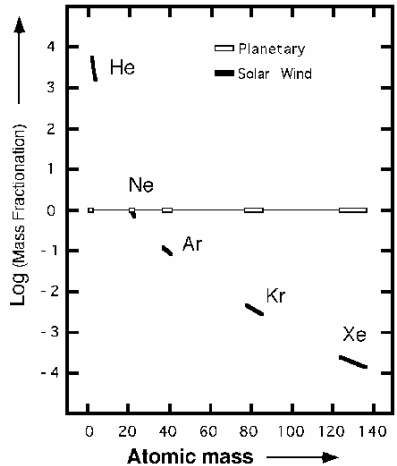

*Figure 5. Excess light isotopes in the solar wind*

Table 1 shows that light isotopes are less abundant in solar flares, as if flares by-pass 3.4 stages of mass fractionation [26]. Heavy elements are enriched systematically, by several orders of magnitude, in material ejected from the interior of the Sun by an impulsive solar flare [27].

Table 1. He, Ne, Mg, and Ar in solar wind and flares

| Isotopic Ratios | Solar Wind | Solar Flares | SW/SF | Expected** |
|---|---|---|---|---|
| $^3$He/$^4$He | 0.00041 | 0.00026 | 1.58 | 1.63 |
| $^{20}$Ne/$^{22}$Ne | 13.6 | 11.6 | 1.17 | 1.18 |
| $^{24}$Mg/$^{26}$Mg | 7.0 | 6.0 | 1.17 | 1.15 |
| $^{36}$Ar/$^{38}$Ar | 5.3 | 4.8 | 1.10 | 1.10 |

**If solar flares by-pass 3.4 stages of fractionation

The probability is almost zero (P < 2 x $10^{-33}$) that Eq. (1) by chance selects from the solar atmosphere seven trace elements that a) all have even atomic numbers, b) are made deep in supernovae, and c) are the same elements that comprise 99% of ordinary meteorites [1].

**E.** Systematic properties [28] of 2,850 known nuclides (Fig. 6) reveal an inherent instability in assemblages of neutrons relative to neutron emission. This may explain luminosity [29-31] of an iron-rich Sun. Neutrons emitted from the collapsed SN core may initiate a chain of reactions that generate luminosity, solar neutrinos, and an outpouring in the solar wind of 3 x $10^{43}$ H$^+$ per year.

- Escape of neutrons from the collapsed solar core
  $<_0^1n> \rightarrow {}_0^1n\ +\ \sim 10\text{-}22$ MeV
- Neutron decay or capture by other nuclides
  $_0^1n \rightarrow\ _1^1H^+ + e^- + \text{anti-}\nu\ + 1$ MeV

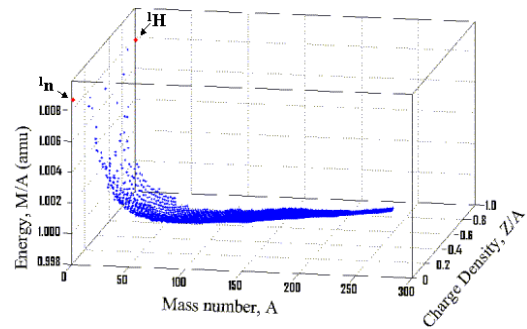

*Figure 6. The cradle of the nuclides.*

- Fusion and upward migration of H$^+$
  $4\ _1^1H^+ + 2\ e^- \rightarrow\ _2^4He^{++} + 2\ \nu\ + 27$ MeV
- Escape of excess H$^+$ in the solar wind
  3 x $10^{43}$ H$^+$/year depart in the solar wind

The hydrogen-filled universe may be the result of this outflow of protons from the Sun and other stars. A summary is on the web <http://www.ballofiron.com>

CONCLUSIONS AND PROPOSED TESTS

The link of Xe-1 with iron extends to the Sun: Iron is its most abundant element. Fusion in the parent star (Fig. 3) likely depleted light elements from the material that formed the Sun and the inner planets.

The following measurements are proposed to test our conclusion of an iron-rich Sun:
1. Measure anti-neutrinos (3 x $10^{38}$ s$^{-1}$, E < 0.782 MeV) from neutron decay at the solar core. Low E targets for inverse β-decay are the Homestake Mine $^{35}$Cl → $^{35}$S reaction [32], the $^{14}$N → $^{14}$C or $^3$He → $^3$H reactions.
2. Measure neutrinos from reactions that increased the $^{15}$N/$^{14}$N ratio [33] and produced excess $^6$Li and $^{10}$Be in the outer layers of the Sun [34,35].
3. Measure microwave background radiation [36] from the supernova explosion here 5 Gy ago.
4. Measure gravity anomalies, magnetic fields, the quadrupole moment, or circular polarized light [37] from a compact object (~10 km) in the Sun.
5. Measure other properties that constrain mass segregation in the Sun and other stars [38-40].
6. Look for excess heavy elements in the fast-moving solar wind, e.g., from the Sun's poles.
7. Use a narrowly focused laser beam to penetrate the Sun's hydrogen-rich veneer.

ACKNOWLEDGEMENT

The Foundation for Chemical Research and the University of Missouri-Rolla supported this research.